\documentclass[12pt,A4]{article}
\usepackage{amssymb, amsmath, enumerate}
\begin{document}
\newtheorem{thm}{Theorem}
\newtheorem{lm}[thm]{Lemma}
\newtheorem{prop}[thm]{Proposition}
\newtheorem{cor}[thm]{Corollary}
\newtheorem{defi}[thm]{Definition}
\newtheorem{eg}[thm]{Example}

\numberwithin{equation}{section}

\newcommand{\proof}{{\bf Proof\ }}
\newcommand{\eproof}{$\blacksquare$\bigskip}
\newcommand{\CC}{\mathcal C}
\newcommand{\CF}{\mathcal F}
\newcommand{\CR}{\mathcal R}
\newcommand{\C}{\Bbb C}
\newcommand{\R}{\Bbb R}

\newcommand{\prob}{\mathop{\mathrm{Pr}}}
\newcommand{\eps}{\varepsilon}
\newcommand{\raction}{\triangleleft}
\newcommand{\Del}{\Delta}
\newcommand{\del}{\delta}
\newcommand{\Lan}{\Lambda}
\newcommand{\Gam}{\Gamma}
\newcommand{\gam}{\gamma}
\newcommand{\ten}{\bigotimes}
\newcommand{\sten}{\otimes}
\newcommand{\Om}{\Omega}
\newcommand{\sfi}{\varphi}
\newcommand{\om}{\omega}
\newcommand{\lan}{\lambda}
\newcommand{\sig}{\sigma}
\newcommand{\al}{\alpha}
\newcommand{\be}{\beta}
\newcommand{\all}{\forall}
\newcommand{\flsh}{\longrightarrow}
\newcommand{\ld}{\ldots}
\newcommand{\cd}{\cdots}
\newcommand{\bu}{\bullet}
\newcommand{\p}{\partial}
\newcommand{\eqn}[2]{\begin{equation}#2\label{#1}\end{equation}}
%\newcommand{\align}\cite{AlbMa}{\begin{eqnarray*}#1\end{eqnarray*}}
% e.g.\align{LHS &&=RHS1\\ &&=RHS2\\ &&=RHS3....}

\textheight 23.6cm \textwidth 16cm \topmargin -.2in \headheight 0in
\headsep 0in \oddsidemargin 0in \evensidemargin 0in \topskip 28pt

\title{Topics on Hamiltonian Mechanics}

\author{S.E.Akrami\footnote{Supported financially by the grant 83810319 from Institute for Research in Fundamental
Sciences, P.O.Box:19395-5746, Tehran, Iran.
E-mail:akramisa@ipm.ir}\\
Departments of Mathematics and Physics,\\ Institute for Research in
Fundamental Sciences,\\ P.O.Box:19395-5746, Tehran, Iran} \maketitle
\begin{abstract}
We modify Hamiltonian mechanics. We reformulate the law of
conservation of energy.
\end{abstract}
\textbf{Acknowledgement} I thank Allah Rabbel-Alamin and Imam Zaman.

This paper is based on the following observation. Consider the
following modified Hamiltonian equations \eqn{c-hamilton}{\frac{\p
Q}{\p t_1}+\frac{\p Q}{\p t_2}=\frac{\p H}{\p P},~~~\frac{\p P}{\p
t_1}+\frac{\p P}{\p t_2}=-\frac{\p H}{\p Q}} for unknown functions
$Q=Q(t_1,t_2),P=P(t_1,t_2)$ and given Hamiltonian $H=H(Q,P)$.

If we set $q(t):=Q(t,t) ,p(t):=P(t,t)$ then since
$\dot{q}(t)=\frac{\p Q}{\p t_1}(t,t)+\frac{\p Q}{\p t_2}(t,t)$ and
$\dot{p}(t)=\frac{\p P}{\p t_1}(t,t)+\frac{\p P}{\p t_2}(t,t)$ we
conclude that $q(t)$ and $p(t)$ satisfy the classical Hamilton's
equations. Thus if we computes the energy $H(q(t),p(t))$ then we get
a constant $H(q(t),p(t))=H(q(0),p(0))$.

But instead of the above method, i.e. before putting $t_1=t_2=t$ if
we expand $Q$ and $P$ in terms of powers of $t_2$ as
\eqn{powers}{Q(t_1,t_2)=\sum_{n=0}^\infty
q_n(t_1)t_2^n,P(t_1,t_2)=\sum_{n=0}^\infty p_n(t_1)t_2^n,} and then
substitute them in the equations (\ref{c-hamilton}) and then by
equating the coefficients of powers of $t_2$ in two sides of each
equation, we get some recursive relations among the unknown
coefficients $q_n(t_1),p_n(t_1)$ which can be solved by knowing the
initial conditions $q_0(t_1),p_0(t_1)$ and then substituting these
coefficients in (\ref{powers}) we obtain $Q=Q(t_1,t_2)$ and
$P=P(t_1,t_2)$ and at last by setting $t_1=t_2=t$ we obtain
$q(t):=Q(t,t) ,p(t):=P(t,t)$. In all examples which we were able to
solve the equations by this method, we observed that  after
calculating the energy $H(q(t),p(t))$ which as mentioned above that
in general one expects to get a constant, in fact we got

\eqn{qconservationlaw}{H(q(t),p(t))=\hat{h}(t),} where
$h(t):=H(q_0(t),p_0(t))$ and
\eqn{alter}{\hat{h}(t):=\sum_{n=0}^{\infty}(-1)^n\frac{h^{(n)}(t)}{n!}t^n.}By
$h^{(n)}$ we mean the $n$-th derivative of $h$.

Since we expect constant energy, the question arises that if for any
function $f(t)$ one has $\hat{f}(t)=f(0)$ for all $t$? In fact if we
differentiate the series term by term we get
\begin{eqnarray}\frac{d}{dt}\hat{f}(t)&=&
\sum_{n=0}^{\infty}(-1)^n\frac{f^{(n+1)}(t)}{n!}t^n+
\sum_{n=1}^{\infty}(-1)^n\frac{f^{(n)}(t)}{(n-1)!}t^{n-1}\nonumber\\&=&
\sum_{n=0}^{\infty}(-1)^n\frac{f^{(n+1)}(t)}{n!}t^n-\sum_{n=0}^{\infty}(-1)^n\frac{f^{(n+1)}(t)}{n!}t^n
\nonumber\\&=&0\nonumber
\end{eqnarray}
Thus $\hat{f}(t)$ should be constant. But the point is that we are
not allowed to differentiate term by term from a series even the
series is uniformly convergent.

We were not able to prove this up to the present time. The only
result is
\begin{thm}
If a function $f$ can be expanded around origin via power series
$f(t)=\sum_{n=0}^{\infty}\frac{f^{(n)}(0)}{n!}t^n,t\in\R$, then
$\hat{f}$ is absolutely and uniformly convergent to the constant
$f(0)$. That is
\eqn{hat0}{\sum_{n=0}^{\infty}(-1)^n\frac{f^{(n)}(t)}{n!}t^n=f(0),~~~t\in\R,}
for analytic functions.
\end{thm}

Another evidence for strangeness of the series $\hat{f}(t)$ comes
from the following argument. Suppose that $f$ is a smooth periodic
function which  by the Fourier analysis we know that it has Fourier
expansion. That is we suppose $f(t)=\sum_{m=-\infty}^\infty
c_me^{im\om t}.$ Then it is well known that we can differentiate to
get $f^{(n)}(t)=\sum(im\om)^n c_me^{im\om t}.$ Thus
\begin{eqnarray}\hat{f}(t)&=&\sum_n\sum_m\frac{(-1)^n(im\om)^{n}}{n!}c_me^{im\om
t}t^n\nonumber\\&=&
\sum_m\sum_n(\frac{(-1)^n(im\om)^{n}}{n!}t^n)c_me^{im\om
t}\nonumber\\&=& \sum_me^{-im\om t}c_me^{im\om
t}\nonumber\\&=&\sum_{m=-\infty}^{\infty}c_m\nonumber\\&=&f(0).\nonumber
\end{eqnarray}
But the above argument is again analytically ill, since we have
exchanged the order of two infinite sums which from the theorems of
mathematical analysis we are not allowed in general to do so. In
order to be able to do so there exists a general theorem on double
series which states that if for a double infinite series
$\sum_{m,n}a_{mn}$ for each $n$ the series $\sum_m |a_{mn}|$ is
convergent which we show its sum by $b_n$ and the series $\sum_n
b_n$ is also convergent then we are allowed to exchange the order of
summation. That is we have $\sum_n\sum_ma_{mn}=\sum_m\sum_na_{mn}$.
Now let us check if this criterion can be applied to the double
series whose entries are $a_{mn}:=\frac{(-im)^n}{n!}c_me^{im t}t^n$.
For simplicity we have assumed that $\om=1$. We have $\sum_m
|a_{mn}|=2\frac{t^n}{n!}\sum_{m=1}^\infty m^n|c_m|$. But in Fourier
analysis it is well known that the series $f^{(n)}(t)=\sum(im)^n
c_me^{im t}$ is absolutely convergent. That is the series
$\sum_{m=1}^\infty m^n|c_m|$ is convergent which we show its sum by
$\al_n$. Thus we have $\sum_m |a_{mn}|=2\frac{\al_nt^n}{n!}.$ Thus
we should verify the convergence of  the series
$\sum_{n=0}^\infty\frac{\al_nt^n}{n!}$. But this is a power series
whose convergence radius is given by $R=\al^{-1}$ where
$\al:=\limsup\sqrt[n]{\frac{\al_n}{n!}}=\limsup\sqrt[n]{\frac{\al_n}{n!}}$.
The point is that we are not sure if $R\ne0$?

In summary the question of convergence of the series $\hat{f}$ and
as well as the constancy of the sum of the series for a periodic or
a general smooth function is remain open.

\textbf{Open Questions} Is it true that for all systems
(\ref{qconservationlaw}) holds?

Is there non-analytic function $f$ such that the series $\hat{f}$ is
point-wise or uniformly convergent and among such functions if there
is any, is there function such that the series converges to a
non-constant function?

We conjecture that there are nonanalytic probably periodic functions
$f$ such that the series $\hat{f}(t)$ is piecewise constant and
takes discrete values. That is $\hat{f}(t)$ is constant only on some
intervals and its values jump when one passes from an interval to
its next interval. More details will appear soon in arXive.

Thus if $q_0(t),p_0(t)$ are simultaneously analytic, that is if both
can be expanded as power series
$$q_0(t)=\sum_{n=0}^{\infty}q_{0n}t^n,~~~p_0(t)=\sum_{n=0}^{\infty}p_{0n}t^n$$
then the function $h(t):=H(q_0(t),p_0(t))$ is also analytic and
therefore we recover the classical conservation law
$H(q(t),p(t))=h(0)=H(q_0(0),p_0(0))=H(q_{00},p_{00})$. Thus we
observe that only the coefficients $q_{00}$ and $p_{00}$ of the
expansion of $q_0(t)$ and $p_0(t)$ contribute in the energy. That is
in the case in which $q_0(t),p_0(t)$ are simultaneously analytic,
dependence of $q_0(t),p_0(t)$ to time does not effect the system and
we can safely, as in the classical mechanics, assume that
$q_0(t)=q_0,p_0(t)=p_0$ are constants. That is from the beginning we
did not need two dimensional time $(t_1,t_2)$ and the modified
equations (\ref{c-hamilton}) are superfluous and the classical
Hamilton's equations are enough. But remember that we were able only
prove the constancy of $\hat{h}(t)$ only for simultaneously analytic
primitive conditions $q_0(t),p_0(t)$. Thus if one of them is not
analytic then we may find new feature in our model. So we need to
study more such a strange object as (\ref{alter})!

The following proposition will be useful in computations.
\begin{thm}
(i)If the series $\hat{f}$ and $\hat{g}$ are convergent then the
series $\widehat{af+bg}$ are also convergent and
$\widehat{af+bg}=a\hat{f}+b\hat{g}$

(ii) If moreover at least one of series $\hat{f}$ and $\hat{g}$ are
absolutely convergent then the series $\widehat{fg}$ is also
convergent and $\widehat{fg}=\hat{f}\hat{g}$.
\end{thm}
\section{Examples}
\begin{eg}$H(q,p)=ap+bq$, where $a$ and $b$
are some given constants.
\end{eg}
\proof The recursive equations obtained from (\ref{c-hamilton})
after expansion (\ref{powers}) are
$(n+1)q_{n+1}+\dot{q}_n=a,(n+1)p_{n+1}+\dot{p}_n=-b.$ One can easily
show by induction that $q_n(t_1)=\frac{(-1)^n}{n!}q^{(n)}_0(t_1),
~~~p_n(t_1)=\frac{(-1)^n}{n!}p^{(n)}_0(t_1)$ for $n\ne1$ and
$q_1(t_1)=a-\dot{q}_0(t_1),p_1(t_1)=\dot{p}_0(t_1)-b.$ Thus
$q(t)=Q(t,t)=\sum_{n=0}^\infty q_n(t)t^n=at+\widehat{q_0}(t)$ and
$p(t)=-bt+\widehat{p_0}(t)$. Finally
$H(q(t),p(t))=-abt+a\widehat{p_0}(t)+bat+b\widehat{q_0}(t)=\widehat{ap_0+bq_0}(t)=\hat{h}(t)$
where $h(t)=ap_0(t)+bq_0(t)=H(q_0(t),p_0(t)).$ \eproof
\begin{eg}The harmonic oscillator
$H(q,p)=\frac{1}{2}p^2+\frac{\om^2}{2}q^2$.
\end{eg}
\proof The recursive equations obtained from (\ref{c-hamilton})
after expansion (\ref{powers}) are
\eqn{h1}{(n+1)q_{n+1}+\dot{q}_n=p_n,~~~(n+1)p_{n+1}+\dot{p}_n=-\om^2
q_n.} We show by induction that for $n>0$
\eqn{h3}{q_n=\frac{(-1)^n}{n!}(\sum_{i=0}^{[\frac{n}{2}]}(-1)^iC^n_{2i}\om^{2i}q_0^{(n-2i)}-\sum_{i=0}^{
[\frac{n-1}{2}]}(-1)^{i}C^n_{2i+1}\om^{2i}p_0^{(n-2i-1)})}and
\eqn{h4}{p_n=\frac{(-1)^n}{n!}(\sum_{i=0}^{[\frac{n}{2}]}(-1)^iC^n_{2i}\om^{2i}p_0^{(n-2i)}+\om^{2}\sum_{i=0}^{
[\frac{n-1}{2}]}(-1)^{i}C^n_{2i+1}\om^{2i}q_0^{(n-2i-1)})} where
$C_i^j:=\frac{j!}{i!(j-i)!}$.

First step of induction $n=1$. We put $n=0$ in (\ref{h1}) to get
$q_1=-\dot{q}_0+p_0$ and $p_1=-\dot{p}_0-\om^2q_0$. On the other
hand if we put $n=1$ in the right hand sides of (\ref{h2}) we get
$q_1=-(\dot{q}_0-p_0)$ and $p_1=-(\dot{p}_0+\om^2 q_0)$. Thus the
first step of induction is verified. Let (\ref{h1}) is true for $n$
we prove it for $n+1$. First suppose $n=2k+1$ is odd. We have
\begin{eqnarray}-\frac{(n+1)!}{(-1)^{n+1}}q_{n+1}&=&\frac{n!}{(-1)^{n}}(-\dot{q}_n+p_n)\nonumber\\&=&
-\sum_{i=0}^{k}(-1)^iC^n_{2i}\om^{2i}q_0^{(n-2i+1)}+\sum_{i=0}^{
k}(-1)^{i}C^n_{2i+1}\om^{2i}p_0^{(n-2i)}\nonumber\\&+&
\sum_{i=0}^{k}(-1)^iC^n_{2i}\om^{2i}p_0^{(n-2i)}+\om^{2}\sum_{i=0}^{
k}(-1)^{i}C^n_{2i+1}\om^{2i}q_0^{(n-2i-1)}\nonumber\\&=&
-\sum_{i=0}^{k}(-1)^iC^n_{2i}\om^{2i}q_0^{(n-2i+1)}+\om^{2}\sum_{i=1}^{
k+1}(-1)^{i-1}C^n_{2(i-1)+1}\om^{2(i-1)}q_0^{(n-2(i-1)-1)}\nonumber\\&+&\sum_{i=0}^{
k}(-1)^{i}C^n_{2i+1}\om^{2i}p_0^{(n-2i)}+
\sum_{i=0}^{k}(-1)^iC^n_{2i}\om^{2i}p_0^{(n-2i)}\nonumber\\&=&
-q_0^{(n+1)}+(-1)^k\om^{2k+2}q_0-\sum_{i=1}^{k}(-1)^i(C^n_{2i}+C^n_{2i-1})\om^{2i}q_0^{(n-2i+1)}\nonumber\\&+&\sum_{i=0}^{
k}(-1)^{i}(C^n_{2i}+C^n_{2i+1})\om^{2i}p_0^{(n-2i)}\nonumber\\&=&
-q_0^{(n+1)}+(-1)^k\om^{2k+2}q_0-\sum_{i=1}^{k}(-1)^iC^{n+1}_{2i}\om^{2i}q_0^{(n-2i+1)}\nonumber\\&+&\sum_{i=0}^{
k}(-1)^{i}C^{n+1}_{2i+1}\om^{2i}p_0^{(n-2i)}\nonumber\\&=&
-\sum_{i=0}^{k+1}(-1)^iC^{n+1}_{2i}\om^{2i}q_0^{(n-2i+1)}+\sum_{i=0}^{
k}(-1)^{i}C^{n+1}_{2i+1}\om^{2i}p_0^{(n-2i)}\nonumber
\end{eqnarray}
Thus
$q_{n+1}=\frac{(-1)^{n+1}}{(n+1)!}(\sum_{i=0}^{[\frac{n+1}{2}]}(-1)^iC^{n+1}_{2i}\om^{2i}q_0^{(n-2i+1)}-\sum_{i=0}^{
[\frac{n}{2}]}(-1)^{i}C^{n+1}_{2i+1}\om^{2i}p_0^{(n-2i)})$. This
means that the part (\ref{h3}) of assertion of induction holds for
$n+1$ if assertion of induction holds for odd $n$. Similarly one can
show that the part (\ref{h4}) of assertion of induction holds for
$n+1$ if assertion of induction holds for odd $n$. The case of even
$n$ is similar.

Moreover we show that if the primitive conditions $q_0(t)$ and
$p_0(t)$ are such that the series $\widehat{q_0}(t)$ and
$\widehat{q_0}(t)$ converges absolutely then
\eqn{h5}{q(t)=\widehat{q_0}(t)\cos\om
t+\om^{-1}\widehat{p_0}(t)\sin\om t} and
\eqn{h6}{p(t)=\widehat{p_0}(t)\cos\om t-\om \widehat{q_0}(t)\sin\om
t.}

Since the series $\cos(\om
t)=\sum_{n=0}^\infty\frac{(-1)^n}{(2n)!}\om^{2n}t^{2n}$ and
$\sin(\om
t)=\sum_{n=0}^\infty\frac{(-1)^n}{(2n+1)!}\om^{2n+1}t^{2n+1}$ are
absolutely convergent,  by the Mertens' theorem \cite{Apo}, we can
multiply the following series by the Cauchy's product rule
\begin{eqnarray}\widehat{q_0}(t)\cos\om t+\om^{-1}\widehat{p_0}(t)\sin\om t&=&
\sum_{n=0}^{\infty}(-1)^n\frac{q_0^{(n)}(t)}{n!}t^n\sum_{n=0}^\infty\frac{(-1)^n}{(2n)!}\om^{2n}t^{2n}\nonumber\\&+&
\om^{-1}\sum_{n=0}^{\infty}(-1)^n\frac{p_0^{(n)}(t)}{n!}t^n\sum_{n=0}^\infty\frac{(-1)^n}{(2n+1)!}\om^{2n+1}t^{2n+1}\nonumber\\&=&
\sum_{n=0}^{\infty}\sum_{i=0}^{[\frac{n}{2}]}(\frac{(-1)^{n-2i}}{(n-2i)!}q_0^{(n-2i)}\frac{(-1)^{i}}{(2i)!}\om^{2i})t^{n-2i}t^{2i}
\nonumber\\&+&\om^{-1}\sum_{n=0}^{\infty}\sum_{i=0}^{[\frac{n-1}{2}]}(\frac{(-1)^{n-2i-1}}{(n-2i-1)!}p_0^{(n-2i-1)}\frac{(-1)^{i}}{(2i+1)!}\om^{2i+1})t^{n-2i-1}t^{2i+1}
\nonumber\\&=&\sum_{n=0}^{\infty}(\frac{(-1)^n}{n!}\sum_{i=0}^{[\frac{n}{2}]}(-1)^iC^n_{2i}\om^{2i}q_0^{(n-2i)}\nonumber\\&-&\sum_{i=0}^{
[\frac{n-1}{2}]}(-1)^{i}C^n_{2i+1}\om^{2i}p_0^{(n-2i-1)}))t^n\nonumber\\&=&\sum_{n=0}^{\infty}q_n(t)t^n\nonumber\\&=&Q(t,t)
\nonumber\\&=&q(t)\nonumber
\end{eqnarray}
This proves  (\ref{h5}). Similarly one can prove (\ref{h6}).

Thus
$H(q(t),p(t))=\frac{1}{2}\widehat{p_0}^2+\frac{\om^2}{2}\widehat{q_0}^2=\hat{h}(t)$,
where
$h(t):=\frac{1}{2}p_0^2(t)+\frac{\om^2}{2}q^2_0(t)=H(q_0(t),p_0(t)).$
\eproof
\begin{eg}
$H(q,p)=pq^2.$
\end{eg}
\proof The recursive equations obtained from (\ref{c-hamilton})
after expansion (\ref{powers}) are
\eqn{}{(n+1)q_{n+1}+\dot{q}_n=\sum_{i=0}^nq_iq_{n-i},~~~(n+1)p_{n+1}+\dot{p}_n=-2\sum_{i=0}^nq_ip_{n-i}.}
Solution of these equations are difficult. So instead of solving
them, we just show that the few first terms of the series
$q(t)=Q(t,t)=\sum_{n=0}^{\infty}q_n(t)t^n$ and
$\frac{\hat{q}_0(t)}{1-t\hat{q}_0(t)}$ coincide  and also the few
first terms of the series $p(t)=P(t,t)=\sum_{n=0}^{\infty}p_n(t)t^n$
and $\hat{p}_0(t)(1-t\hat{q}_0(t))^2$ coincide. That is we
conjecture that
\eqn{x}{q(t)=\frac{\hat{q}_0(t)}{1-t\hat{q}_0(t)},~~~p(t)=\hat{p}_0(t)(1-t\hat{q}_0(t))^2.}
By solving the first few terms of these recursive relations we get
$q_1=q_0^2-q_0',q_2=q_0^3,-2q_0q_0'+\frac{1}{2}q_0'',q_3=q_0^4-3q_0^2q_0'+q_0q_0''+q_0'^2-\frac{1}{6}q_0''',\ld$

On the other hand if we set $\sum_{n=0}^\infty
a_nt^n:=\frac{\dot{q}_0(t)}{1-t\dot{q}_0(t)}$. Then
\begin{eqnarray}\sum_{n=0}^\infty\frac{(-1)^n}{n!}q_0^{(n)}t^{n}&=&
(1-\sum_{n=0}^\infty\frac{(-1)^n}{n!}q_0^{(n)}t^{n+1})\sum_{n=0}^\infty
a_nt^n\nonumber\\&=&a_0+(a_1-a_0q_0)t+(a_2-a_1q_0+a_0q_0')t^2\nonumber\\&+&(a_3-a_2q_0+a_1q_0'-\frac{1}{2}a_0q_0'')t^3+\cd\nonumber
\end{eqnarray}Thus
$a_0=q_0,a_1-a_0q_0=-q_0',a_2-a_1q_0+a_0q_0'=\frac{1}{2}q_0'',a_3-a_2q_0+a_1q_0'-\frac{1}{2}a_0q_0''=\frac{-1}{6}q_0''',\ld$.
Thus
$a_0=q_0,a_1=q_0^2-q_0',a_2=q_0^3,-2q_0q_0'+\frac{1}{2}q_0'',a_3=q_0^4-3q_0^2q_0'+q_0q_0''+q_0'^2-\frac{1}{6}q_0''',\ld$.Hence
we see that $a_0=q_0,a_1=q_1,a_2=q_2,a_3=q_3,\ld$. This proves the
first relation of (\ref{x}). The other part is proved similarly.

Up to the present time we have not been able to prove the conjecture
(\ref{x}). But suppose this conjecture is true. Then we have
$H(q(t),p(t))=(\dot{q}_0(t))^2\dot{p}_0(t)=\hat{h}(t)$ where
$h(t)=(q_0(t))^2p_0(t)=H(q_0(t),p_0(t))$. \eproof


\begin{thebibliography}{}
\bibitem{Apo}
Tom M. Apostol \emph{Mathematical Analysis}, Addison-Wesley,1975
\end{thebibliography}
\end{document}